\documentclass[entropy,article,accept,moreauthors,10pt,a4paper]{mdpi}

\firstpage{1}
\pubvolume{27}
\issuenum{1}
\articlenumber{95}
\pubyear{2025}
\copyrightyear{2025}
 \externaleditor{Academic Editor: Santi Prestipino} 
\datereceived{2 December 2024}
\daterevised{15 January 2025} 
\dateaccepted{17 January 2025}
\datepublished{20 January 2025}
\hreflink{https://doi.org/10.3390/e27010095} 

\newcommand{\dd}{\mathrm{d}}
\newcommand{\bp}{\beta p}
\newcommand{\la}{\lambda}
\newcommand{\e}{\eta}

\def\one{{(1)}}
\def\two{{(2)}}
\def\three{{(3)}}
\def\four{{(4)}}

\newcommand{\hr}{{\text{HR}}}

\newcommand\beq{\begin{equation}}
\newcommand\eeq{\end{equation}}
\newcommand\beqa{\begin{eqnarray}}
\newcommand\eeqa{\end{eqnarray}}

\def\bal#1\eal{\begin{align}#1\end{align}}

\Title{{Discontinuous} 
 Structural Transitions in Fluids with \mbox{Competing Interactions}}

\TitleCitation{Discontinuous Structural Transitions in Fluids with Competing Interactions}

\Author{{Ana M. Montero} 
 $^{1}$\orcidA{}, {Santos B. Yuste} 
 $^{1,2}$\orcidB{}, Andr\'es Santos $^{1,2,}$*\orcidC{} and Mariano L\'opez de Haro $^{3}$\orcidD{}}

\AuthorNames{Ana M. Montero, Santos B. Yuste, Andr\'es Santos and Mariano L\'opez de Haro}

\AuthorCitation{Montero, A.M.; Yuste, S.B.; Santos, A.; de Haro, M.L.}

\address{%
$^{1}$ \quad Departamento de F\'isica, Universidad de Extremadura, {E-06006} 
Badajoz, Spain; anamontero@unex.es; santos@unex.es\\
$^{2}$ \quad Instituto de Computaci\'on Cient\'ifica Avanzada (ICCAEx), Universidad de
Extremadura, E-06006 Badajoz, Spain\\
 $^{3}$ \quad Instituto de Energ\'ias Renovables, Universidad Nacional Aut\'onoma de M\'exico {(UNAM),} 
 \linebreak Temixco {62580}, Mexico; malopez@unam.mx}

\corres{Correspondence: andres@unex.es}

\abstract{
This paper explores how competing interactions in the intermolecular potential of fluids affect their structural transitions. This study employs a versatile potential model with a hard core followed by two constant steps, representing wells or shoulders, analyzed in both one-dimensional (1D) and three-dimensional (3D) systems. Comparing these dimensionalities highlights the effect of confinement on structural transitions.
Exact results are derived for 1D systems, while the rational function approximation is used for unconfined 3D fluids. Both scenarios confirm that when the steps are repulsive, the wavelength of the oscillatory decay of the total correlation function evolves with temperature either continuously or discontinuously. In the latter case, a discontinuous oscillation crossover line emerges in the temperature--density plane. For an attractive first step and a repulsive second step, a Fisher--Widom line appears.
Although the 1D and 3D results share common features, dimensionality introduces differences: these behaviors occur in distinct temperature ranges, require deeper wells, or become attenuated in 3D. Certain features observed in 1D may vanish in 3D.
We conclude that fluids with competing interactions exhibit a rich and intricate pattern of structural transitions, demonstrating the significant influence of dimensionality and interaction features.}


\begin{document}


\section{Introduction}
\label{sec1}

It is well known that both statistical mechanics and thermodynamics aim at explaining the same phenomena concerning, among other issues, energy, work, and heat exchange in different systems. While the first approach involves a purely microscopic approximation, the second one is macroscopic in nature. Nevertheless, one of the major purposes of statistical physics is the interpretation and prediction of the macroscopic properties of a system in terms of the interactions between its particles. In the case of liquids, one attempts to understand why and under what circumstances certain phases are stable in well-defined intervals of density and temperature and also to try to relate the thermodynamic, structural, and dynamic properties of those phases with the form and size of the molecules that form the liquid and the nature of the intermolecular interactions \cite{M76}.

For the description of a multibody system such as a liquid, it is often enough to consider simplified representations which are able to capture the essential elements of real interactions and lead to an adequate description of the observed phenomenology. Therefore, the great attention that has been paid during many decades to interaction potentials consisting of a hard core followed by one or many piecewise constant sections of different widths and heights (which include the square-well and the square-shoulder potentials) is not surprising \cite{CMA84,CSD89,BG99,FMSBS01,MFPSBS02,SBFMS04,MFSBS05,BPGS06,CBR07,GCC07,RPPS08,OFNB08,ONB09,RPSP10,HTS11,YSH11,BOO11,LC12,SYH12,BOO13,SYHBO13,KK13,YSH22,PMPSGLVTC22,LS24}. With this class of potentials, it has been possible to model and understand many phenomena, such as liquid--liquid transitions \cite{SBFMS04,MFSBS05,CBR07}, colloidal interactions \cite{GCC07}, the density anomaly in water and supercooled liquids \cite{OFNB08,ONB09}, and the thermodynamic and transport properties of Lennard--Jones fluids \cite{CMA84,CSD89}. In particular, in the case of colloidal dispersions, the interaction between a pair of macromolecules is modeled through an effective potential with a short-range attractive part and a long-range repulsive part \cite{RZ21,KPHC22,CZLEM23}. The competition between both parts of this potential leads to an interesting phenomenology and induces changes in phase behavior and in the thermodynamic, structural, and transport properties of the system \cite{PMPSGLVTC22,TCDN24}. Similarly, in the case of complex fluids, such competing interactions are associated with the aggregation or clustering of surfactants, macromolecules,  and colloidal particles in solution, which in turn may produce self-assembly and microphase segregation \cite{IR06,AW07,BBCH12,BT16a,BT16b,HC18,MS20b,BCB20,GMMM22,MCMBP22,MPBC22,CMBMPP23,BPCMMP24}.

There is an extensive body of research on the thermodynamic and structural properties of fluids whose molecules interact via competing attractive and repulsive forces. Particular attention has been given to systems described by variants of the short-range attraction and long-range repulsion (SALR) potential. These include models such as the two-Kac potential, the double Yukawa potential, the Lennard-Jones potential followed by a repulsive Yukawa tail, and the square-well potential followed by a repulsive ramp. For examples and further details, see Refs.~\cite{DL97,SG99,PJPR00,MSTZ04,SMZT04,APER07,AW07,AIPR08,BC12,LC12,PCA13,KK14,GVCWL14,SFL14,CDBC15,ZC16,LX19,B19,CK20,RZ21, VLC21,CMBMPP23}.

In colloidal systems, the competition between short-range attraction and long-range repulsion leads to the emergence of intermediate-range-order structures, resulting in the formation of stable periodic microphases. This competition also disrupts the liquid--vapor phase transition, with the specific form of the SALR potential significantly influencing the morphology of the resulting structures \cite{LC12, LX19}. The intermediate-range order is closely linked to a peak in the static structure factor $S(k)$ (where $k$ is the wavenumber). Specifically, a divergence of $S(k)$ at $k = 0$ indicates an instability associated with large-scale density fluctuations, while a divergence at a finite wavenumber signifies the presence of periodic~microphases.

Despite the extensive research on fluids whose molecules interact via competing attractive and repulsive forces, certain aspects of structural transitions in these systems remain unexplored. The decay of the total correlation function, $h(r) = g(r) - 1$, where $g(r)$ is the radial distribution function, serves as a key indicator of such transitions. These transitions, characterized by oscillatory or monotonic decay, reflect changes in the spatial arrangement of particles arising from the delicate balance between attraction and repulsion in the intermolecular potential. A deeper understanding of the decay behavior of $h(r)$ is crucial for unraveling phenomena such as crystallization, phase separation, self-assembly, and the mechanical properties of complex materials.

All this serves as a motivation for the present paper.
In previous work, we have used the so-called rational function approximation (RFA) approach \cite{HYS08,S16}
to study various three-dimensional fluids whose intermolecular potentials consist of a hard core followed by piecewise constant sections \cite{YSH11,SYH12,SYHBO13}. This includes not only square-well and square-shoulder fluids but also systems where the intermolecular potential combines square shoulders and square wells \cite{YSH22}.
We have also carried out studies of the asymptotic behavior of the direct and total  correlation functions of binary hard-sphere fluid mixtures \cite{PBYSH20,PYSHB21}, which, among other things, exhibit interesting phenomenology concerning structural transitions.

In this paper, we aim to illustrate the effect of competing interactions on structural transitions in fluids. To this end, we consider a fluid of the number density $\rho$ and absolute temperature $T$, where the intermolecular pair potential is given {by} 

\begin{equation}
\varphi(r)=\left\{
\begin{array}{ll}
\infty  ,& r<\sigma, \\
\epsilon_1  ,& \sigma < r <\lambda_1 \sigma,\\
\epsilon_2  ,& \lambda_1\sigma<r<\lambda_2 \sigma, \\
0,&r>\lambda_2 \sigma.
\end{array}
\right.
\label{varphi}
\end{equation}
This potential includes a hard core of the diameter $\sigma$ and two steps characterized by the heights $\epsilon_1$ and $\epsilon_2$ and widths $(\lambda_1-1)\sigma$ and $(\lambda_2-\lambda_1)\sigma$, respectively.
The parameters $\lambda_1$ and $\lambda_2$ are constants satisfying $1 < \lambda_1 < \lambda_2$, where $\lambda_2\sigma$ denotes the total range of the potential.
The sign of each $\epsilon_j$ ($j=1,2$) determines whether the corresponding step is a shoulder ($\epsilon_j > 0$) or a well ($\epsilon_j < 0$).
This form of the potential is flexible enough to explore various competing interactions. In particular, when $\epsilon_1 = \epsilon_2$ or $\epsilon_2 = 0$, the potential reduces to either the square-shoulder potential (for $\epsilon_1 > 0$) or the square-well potential (for $\epsilon_1 < 0$), making these cases particular limits of the general model.
Studies on certain thermodynamic and structural properties of fluids whose molecules interact via a potential of the form given in Equation~\eqref{varphi} have been reported in Refs.~\cite{DL97, LC12, KK13}. However, it is important to note that, in our case, the range of the repulsive interaction is relatively short and cannot be accurately described as long-range.

This work focuses on examining the qualitative changes in the structural behavior of a system as the potential transitions from the square-shoulder case to more complex potentials, where the second section is always a repulsive barrier ($\epsilon_2 > 0$).

If both $\epsilon_1$ and $\epsilon_2$ are positive, the total correlation function $h(r)$ is expected to exhibit oscillatory decay. At very low temperatures, this decay has a wavelength in the order of the range of the repulsive barrier ($\lambda_2 \sigma$). Conversely, at very high temperatures, the wavelength aligns with the hard core diameter ($\sigma$). At a given density, the transition between these behaviors can occur either continuously or discontinuously. In the latter scenario, a discontinuous oscillation crossover (DOC) line would emerge, akin to the one observed in binary hard-sphere mixtures \cite{PBYSH20, PYSHB21, GDER04, GDER05, SPTER16, RCDRSSV24}.

On the other hand, if $\epsilon_1 < 0$ and $\epsilon_2 > 0$, one might expect the presence of a Fisher--Widom (FW) line, which separates a region in the $T$ vs. $\rho$ plane where the asymptotic decay of $h(r)$ is damped in an oscillatory way 
from a region where the decay is purely exponential and monotonic. For a given $\epsilon_2 > 0$, a competition between a DOC line and an FW line could arise as $\epsilon_1$ transitions from positive to increasingly negative values.

From this point onward, we adopt the hard core diameter as the unit of length ($\sigma = 1$), so all distances will be expressed in units of $\sigma$. The reduced density is then given by $\rho^* = \rho \sigma^d$, where $d$ is the dimensionality of the system. Since we assume $\epsilon_2 > 0$ throughout, we use $\epsilon_2$ as the unit of energy and define the reduced temperature as $T^* = k_B T / \epsilon_2$, with $k_B$ being the Boltzmann constant. However, when analyzing the impact of the second barrier on the FW line (in cases where $\epsilon_1 < 0$), we also introduce a second reduced temperature, $T_1^* = k_B T / |\epsilon_1| = T^* \epsilon_2 / |\epsilon_1|$, to capture the relevant energy scale. The key dimensionless parameters characterizing the potential are thus $\lambda_1$, $\lambda_2$, and the ratio $\epsilon_1 / \epsilon_2$.

For reasons that will become apparent later, we restrict the value of $\lambda_2$ to be less than or equal to 2. For symmetry considerations, we generally fix $\lambda_1 = 1.35$ and $\lambda_2 = 1.7$, except in cases where $\epsilon_1 = \epsilon_2$, where the effect of $\lambda_2$ on the DOC line is specifically~examined.

Finally, we note that both one-dimensional (1D) and three-dimensional (3D) fluids interacting via the potential $\varphi(r)$, as defined in Equation (\ref{varphi}), will be examined in the following analysis. This dual approach allows us to explore the impact of strong confinement on structural transitions in fluids with competing interaction potentials. The results for the 1D system will be derived from the exact general solution, while for the unconfined 3D system, we will employ the RFA.

The paper is organized as follows. In Section \ref{sec2}, we consider a 1D fluid. This is followed in Section \ref{sec3} by the parallel analysis of an unconfined 3D fluid, where a brief but self-contained description of the RFA method is provided. Section \ref{sec4} concludes the paper with a discussion of the results, including the differences in the structural behavior of 1D and 3D fluids modeled with the same interaction potential, along with some concluding remarks. Mathematical details are presented in {Appendices} 
 \ref{appA} and \ref{appB}.

\section{The 1D System: Exact Results}
\label{sec2}
\subsection{Theoretical Background}\label{sec2.1}
We begin by considering a system confined to a 1D geometry. In this case, we can take advantage of the fact that Equation \eqref{varphi} satisfies the conditions that, for 1D fluids, lead to exact results for thermodynamic and structural properties \cite{S16}, namely that $\lim_{r \to 0}\varphi(r) = \infty$ and $\lim_{r \to \infty}\varphi(r) = 0$ and that each particle interacts only with its two nearest neighbors when $\lambda_2 \leq 2$.

As in previous works on 1D fluids \cite{SZK53, LZ71, P76, P82, HC04, MS19, MRYSH24}, it is convenient to work with the Laplace transforms of both the radial distribution function $g(r)$ and the Boltzmann factor $e^{-\beta \varphi(r)}$ (where $\beta \equiv 1/k_BT$). These transforms are, respectively, defined as
\beq
G(s)=\int_0^\infty \dd r \, e^{-rs} g(r),\quad\Omega(s)=\int_0^\infty \dd r\, e^{-rs} e^{-\beta \varphi(r)} .
\eeq
In fact, working in the isothermal--isobaric ensemble, one can express $G(s)$ in terms of $\Omega(s)$ as \cite{S16}
\begin{equation}
\label{G(s)}
G(s)=\frac{\Omega'(\bp)}{\Omega(\bp)}\frac{\Omega(s+\bp)}{\Omega(s+\bp)-\Omega(\bp)}\,,
\end{equation}
where $p$ is the pressure and $\Omega'(s)\equiv \partial_s\Omega(s)=-\int_0^\infty \dd r\, e^{-rs} r e^{-\beta \varphi(r)}$.
Furthermore, the density of the fluid is also related to $\Omega(s)$ and reads as
\beq
\label{Z}
\rho=-\frac{\Omega(\bp)}{\Omega'(\bp)}.
\eeq

In principle, the total correlation function $h(r)$ can be expressed in terms of the infinite set of poles $\{s_n\}$ of $G(s)$, which correspond to the nonzero roots of $\Omega(s + \bp) = \Omega(\bp)$. These poles have negative real parts and may be either real ($s_n = -\kappa_n$) or form complex--conjugate pairs ($s_n = -\zeta_n \pm \imath \omega_n$). For simplicity, we will use the term ``pole'' to refer collectively to both real values and complex--conjugate pairs. The locations of these poles depend on the thermodynamic state, with the pole whose real part is closest to zero governing the asymptotic behavior of the total correlation function.

In the case where the leading and subleading poles (i.e., the two poles with real parts closest to zero) are both complex ($s_1 = -\zeta_1 \pm \imath\omega_1$ and $s_2 = -\zeta_2 \pm \imath\omega_2$), one has
\begin{equation}
\label{h(r)_1D_1}
h(r)\approx
2|A_{\zeta_1}|e^{-\zeta_1 r}\cos(\omega_1 r+\delta_1)+
2|A_{\zeta_2}|e^{-\zeta_2 r}\cos(\omega_2 r+\delta_2), \quad r\gg 1,
\end{equation}
where $\delta_n$ is the argument of the associated residue $|A_{\zeta_n}|e^{\pm \imath\delta_n}$. The first term on the right-hand side of Equation \eqref{h(r)_1D_1} dominates over the second one if $\zeta_1 < \zeta_2$; conversely, the second term dominates if $\zeta_1 > \zeta_2$.
Given a value of $\bp$, there may exist a certain temperature at which the conditions $\zeta_1 = \zeta_2$ and $\omega_1 \neq \omega_2$ are satisfied. The set of such states plotted on the $T$ vs. $\bp$ plane (or equivalently on the $T$ vs. $\rho$ plane) defines the DOC line. When this line is crossed, the wavelength of the asymptotic damped oscillations in $h(r)$ undergoes a discontinuous shift from $2\pi/\omega_1$ to $2\pi/\omega_2$ (or vice versa).

Analogously, if the leading and subleading poles  consist of a pair of complex conjugates ($s_1 = -\zeta \pm \imath\omega$) and a real value ($s_2 = -\kappa$), one has
\begin{equation}
\label{h(r)_1D}
h(r)\approx
2|A_\zeta|e^{-\zeta r}\cos(\omega r+\delta)+
A_\kappa e^{-\kappa r}, \quad r\gg 1.
\end{equation}
For a given value of $\bp$, there may exist a specific temperature at which the conditions $\zeta = \kappa$ and $\omega \neq 0$ are satisfied. The collection of such states, when plotted on the $T$ vs. $\bp$ plane (or equivalently on the $T$ vs. $\rho$ plane), defines the FW line. Upon crossing this line, the nature of the decay of the total correlation function $h(r)$ transitions between damped oscillatory and monotonic behavior (or vice versa).

As shown in Appendix \ref{appA1}, all nonzero poles of $G(s)$ are complex if $\varphi(r) \geq 0$, ruling out the possibility of an FW line in such cases. This result applies to the double-step potential given by Equation \eqref{varphi} when $\epsilon_j \geq 0$, including the case $0 \leq \epsilon_1 < \epsilon_2$, where the interaction is effectively attractive within the range $1 < r < \lambda_1$.

For the potential given in Equation (\ref{varphi}), the expressions for $\Omega(s)$  and $\Omega'(s)$ are
\begin{subequations}
\label{Omega}
\beq
\label{Omega(s)}
\Omega(s)=\frac{E_0(s)+E_1(s)+E_2(s)}{s},
\eeq
\beq
\label{Omega'(s)}
\Omega'(s)=-\frac{\Omega(s)}{s}-\frac{E_0(s)+\lambda_1 E_1(s)+\lambda_2E_2(s)}{s},
\eeq
\end{subequations}
where we have set $\sigma = 1$ and introduced the shorthand notation
\beq
E_0(s)\equiv e^{-\beta\epsilon_1}e^{-s},\quad E_1(s)\equiv \left(e^{-\beta\epsilon_2}-e^{-\beta\epsilon_1}\right)e^{-\lambda_1 s},\quad E_2(s)\equiv\left(1-e^{-\beta\epsilon_2}\right)e^{-\lambda_2 s}.
\eeq

Thus, the density, as a function of pressure and temperature,  is given by
\beq
\label{EOS}
\rho^*=\left[\frac{1}{\bp}+\frac{E_0(\bp)+\lambda_1E_1(\bp)
+\lambda_2E_2(\bp)}{E_0(\bp)+E_1(\bp)+E_2(\bp)}\right]^{-1}.
\eeq

The real and imaginary parts of the complex poles of $G(s)$ are the solutions {to} 
\begin{subequations}
\label{eq:poles}
\beq
\label{eq:poles_a}
1-\frac{\zeta}{\bp}=\frac{E_0(\bp) e^{\zeta}\cos\omega+E_1(\bp)e^{\zeta\lambda_1}
\cos(\omega\lambda_1)+E_2(\bp)e^{\zeta\lambda_2}\cos(\omega\lambda_2)}{E_0(\bp)+E_1(\bp)+E_2(\bp)},
\eeq
\beq
\label{eq:poles_b}
-\frac{\omega}{\bp}=\frac{E_0(\bp) e^{\zeta}\sin\omega+E_1(\bp)e^{\zeta\lambda_1}
\sin(\omega\lambda_1)+E_2(\bp)e^{\zeta\lambda_2}\sin(\omega\lambda_2)}{E_0(\bp)+E_1(\bp)+E_2(\bp)}.
\eeq
\end{subequations}

Regardless of the sign of $\epsilon_j$, the leading pole at a given density, $\rho^*$, for the high-temperature limit ($\beta \to 0$) is given by $\zeta = \zeta_\hr(\rho^*)$ and $\omega = \omega_\hr(\rho^*)$, as shown in Appendix \ref{appA2}, where the subscript HR refers to the hard rod fluid. The HR oscillation frequency satisfies $\frac{1}{2} < \omega_\hr(\rho^*)/2\pi < 1$, with the lower and upper bounds corresponding to $\rho^* \to 0$ and $\rho^* \to 1$, respectively. On the other hand, if $\epsilon_j > 0$ and $\rho^* < \lambda_2^{-1}$, the leading pole for the low-temperature limit ($\beta \to \infty$) is given by $\zeta = \lambda_2^{-1} \zeta_\hr(\rho^* \lambda_2)$ and $\omega = \lambda_2^{-1} \omega_\hr(\rho^* \lambda_2)$ (see Appendix \ref{appA3.1}). However, the low-temperature limit for $\lambda_2^{-1} < \rho^* < 1$ is more intricate, as detailed in Appendix \ref{appA3.2}.

If $\epsilon_1<0$ and real poles do exist, they are the solutions to
\beq
\label{eq:poles_c}
1-\frac{\kappa}{\bp}=\frac{E_0(\bp) e^{\kappa}+E_1(\bp)e^{\kappa\lambda_1}
+E_2(\bp)e^{\kappa\lambda_2}}{E_0(\bp)+E_1(\bp)+E_2(\bp)}.
\eeq

\subsection{$\epsilon_1=\epsilon_2>0$:   Influence of $\lambda_2$ on DOC Line}\label{sec2.2}

In the case $\epsilon_1 = \epsilon_2 > 0$, the potential in Equation \eqref{varphi} simplifies to a hard core plus a square shoulder of the width $\lambda_2 - 1$.

\begin{figure}[H]
\begin{adjustwidth}{-\extralength}{0cm}
\centering
\includegraphics[width=15.5cm]{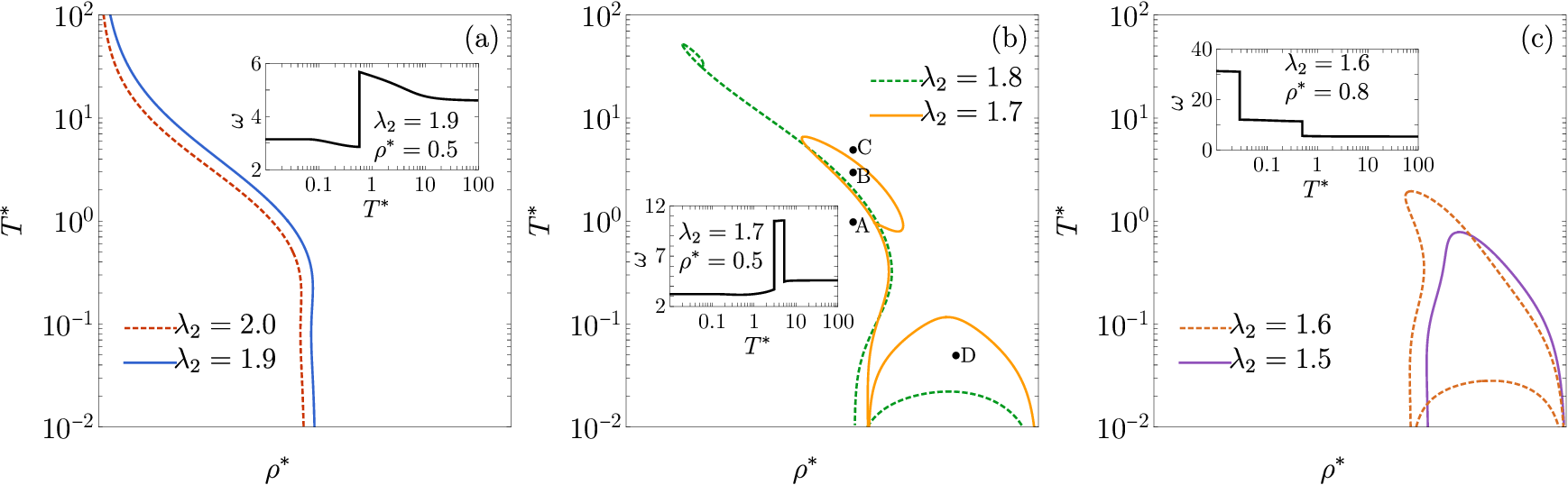}
\end{adjustwidth}

\vspace{-3pt}
\caption{
{DOC} 
 lines on the $T^*$ vs. $\rho^*$ plane for the 1D case with $\epsilon_1=\epsilon_2$: (\textbf{a}) $\lambda_2=2,1.9$, \mbox{(\textbf{b}) $\lambda_2=1.8,1.7$}, and (\textbf{c}) $\lambda_2=1.6,1.5$. Insets display the angular frequency of the asymptotic oscillations of $h(r)$ as a function of $T^*$ for (\textbf{a}) $\lambda_2=1.9$ at $\rho^*=0.5$, (\textbf{b}) $\lambda_2=1.7$ at $\rho^*=0.5$, and (\textbf{c}) $\lambda_2=1.6$~at $\rho^*=0.8$. The circles in panel (\textbf{b}) represent the four states examined in Figure \ref{fig2} for $\lambda_2=1.7$. \label{fig1}}
\end{figure}

As shown in Figure \ref{fig1}, the DOC line exhibits an intricate behavior as $\lambda_2$ varies. For $\lambda_2 = 2$ and $\lambda_2 = 1.9$, distinct DOC lines emerge, each starting at $\rho^* = \lambda_2^{-1}$ for the low-temperature region and shifting toward lower densities as the temperature increases.
When $\lambda_2 = 1.8$, the DOC line intersects with a DOC loop at $\rho^* \approx 0.18$ and $T^* \approx 35$. Inside the loop, the oscillation frequency reaches the values $\omega/2\pi\approx 3/\lambda_2$, significantly larger than outside the loop. The intersection between the DOC line and the DOC loop acts as a triple point, where three distinct complex poles share the same real part, $\zeta$. An additional DOC arc appears, extending between $\rho^* = \lambda_2^{-1}$ and $\rho^* = 1$ for the low-temperature region, within which $\omega$ reaches even higher values ($\omega/2\pi\approx 5=9/\lambda_2$; see Appendix \ref{appA3.2}) than inside the DOC loop.
For $\lambda_2 = 1.7$, the loop expands, shifting toward higher densities and lower temperatures, while the  DOC arc  broadens. Below $T^*=0.01$ (not shown in the figure), an inner arc emerges, which is absent in the case $\lambda_2=1.8$. In the region between the inner and outer arcs for $\lambda_2=1.7$, $\omega/2\pi\approx 5/\lambda_2$, whereas $\omega/2\pi\approx 10=17/\lambda_2$ within the inner arc. As $\lambda_2$ further decreases to $1.6$, the original DOC line vanishes, with the loop and outer arc merging into a more complex DOC region (where $\omega/2\pi\approx 3/\lambda_2$) and the inner arc region (where $\omega/2\pi\approx 5=8/\lambda_2$) growing. At $\lambda_2=1.5$, only the inner arc persists, with $\omega/2\pi\approx 2=3\lambda_2$ within. This evolution illustrates an increasingly complex pattern of structural transitions as the DOC line transforms with decreasing $\lambda_2$.

The insets in Figure \ref{fig1} illustrate the temperature dependence of $\omega$ at several densities and values of $\lambda_2$. In the insets of Figure \ref{fig1}{a},{b}, $\omega$ transitions from $\lambda_2^{-1}\omega_\hr(\rho^*\lambda_2)$ at a low $T^*$ to $\omega_\hr(\rho^*)$ at a high $T^*$. In the inset of Figure \ref{fig1}{a}, a single discontinuous shift is observed as the DOC line is traversed. However, in the inset of Figure \ref{fig1}{b}, two distinct discontinuous jumps in $\omega$ occur as the DOC loop is crossed. The inset of Figure \ref{fig1}{c} shows two discontinuous drops in $\omega$ when crossing the  DOC's inner and outer arcs.

\begin{figure}[H]
\begin{adjustwidth}{-\extralength}{0cm}
\centering
\includegraphics[height=5cm]{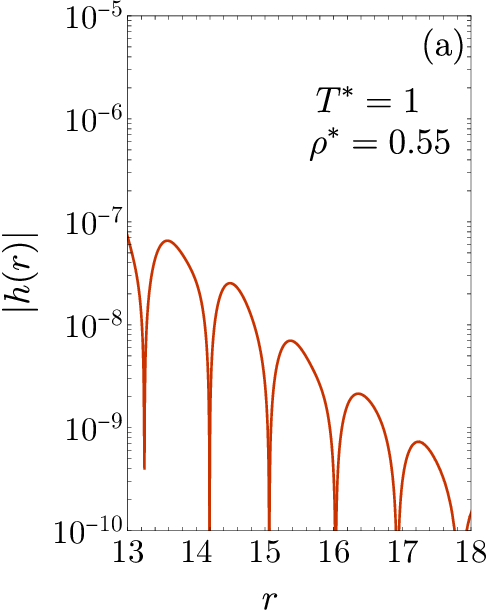}\includegraphics[height=5cm]{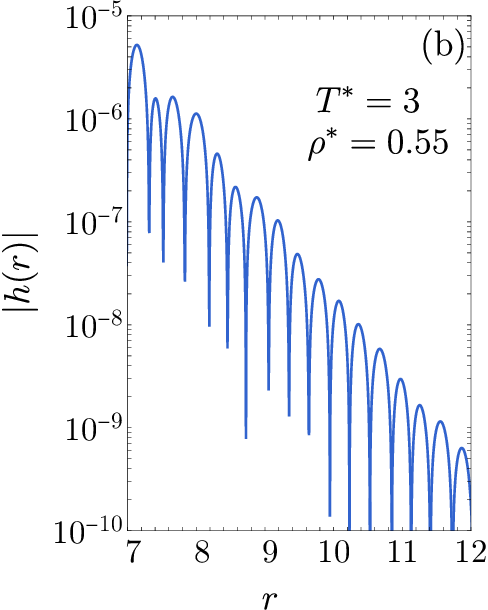}\includegraphics[height=5cm]{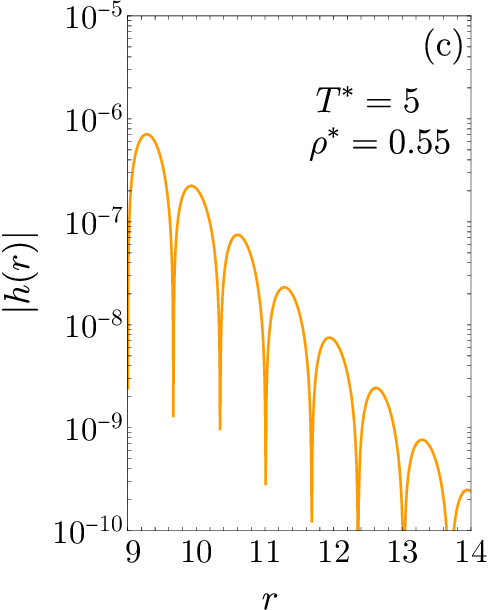}\includegraphics[height=5cm]{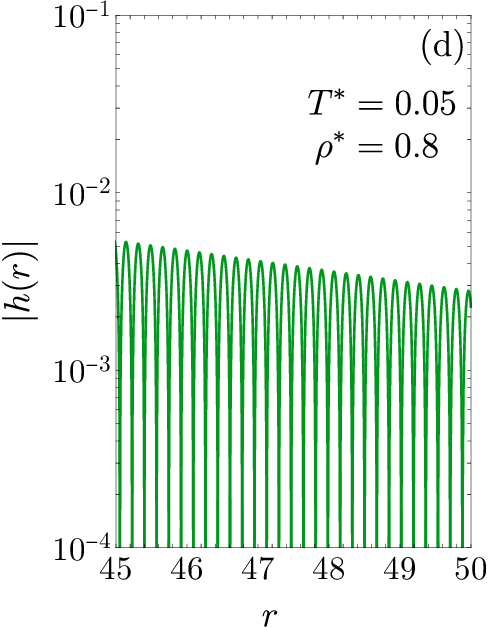}
\end{adjustwidth}

\vspace{-3pt}
\caption{A logarithmic plot of $|h(r)|$ for large $r$ in the 1D case $\epsilon_1=\epsilon_2$, $\lambda_2=1.7$,  for the following states: (\textbf{a}) $(\rho^*,T^*)=(0.55,1)$, (\textbf{b}) $(\rho^*,T^*)=(0.55,3)$, (\textbf{c}) $(\rho^*,T^*)=(0.55,5)$, and (\textbf{d}) $(\rho^*,T^*)=(0.8,0.05)$. These states are labeled A--D in Figure \ref{fig1}{b}, respectively.\label{fig2}}
\end{figure}

As further confirmation of the results presented in Figure \ref{fig1}, we numerically invert the Laplace transform given by Equation \eqref{G(s)} using the method described in \cite{EulerILT} to obtain $h(r)$. The results for $\lambda_2 = 1.7$ and four representative states are shown in Figure \ref{fig2}. In Figure \ref{fig2}{a}--{c}, we fix the density $\rho^*$ and examine a temperature ($T^*=1$) below the loop, a temperature ($T^*=3$) inside the loop, and a temperature ($T^*=5$) above the loop. These three states are labeled A--C in Figure \ref{fig1}{b}, respectively. The corresponding leading poles are $(\zeta,\omega)=(1.254, 3.400)$, $(\zeta,\omega)=(1.881,10.564)$, and $(\zeta,\omega)=(1.696, 4.687)$, respectively, which align fully with the damped oscillatory behavior observed in Figure \ref{fig2}{a}--{c}. As a representative state located between the inner and outer arcs, we select $\rho^* = 0.8$ and $T^* = 0.05$ [see label D in Figure \ref{fig1}{b}]. The corresponding values of the decay parameters are $(\zeta, \omega) = (0.131, 18.233)$, as shown in Figure \ref{fig2}{d}.

\subsection{$\lambda_1=1.35$ and $\lambda_2=1.7$: Influence of $\epsilon_1/\epsilon_2$ on DOC Line}\label{sec2.3}

If $\epsilon_1 \neq \epsilon_2$, both $\lambda_1$ and $\lambda_2$ become relevant parameters.  For symmetry reasons, we choose $\lambda_1-1=\lambda_2-\lambda_1$ so that both sections have the same width. As mentioned in Section \ref{sec1} and to maintain concreteness, we henceforth set $\lambda_1 = 1.35$ and $\lambda_2 = 1.7$.

\begin{figure}[H]
\begin{adjustwidth}{-\extralength}{0cm}
\centering
\includegraphics[width=15.5cm]{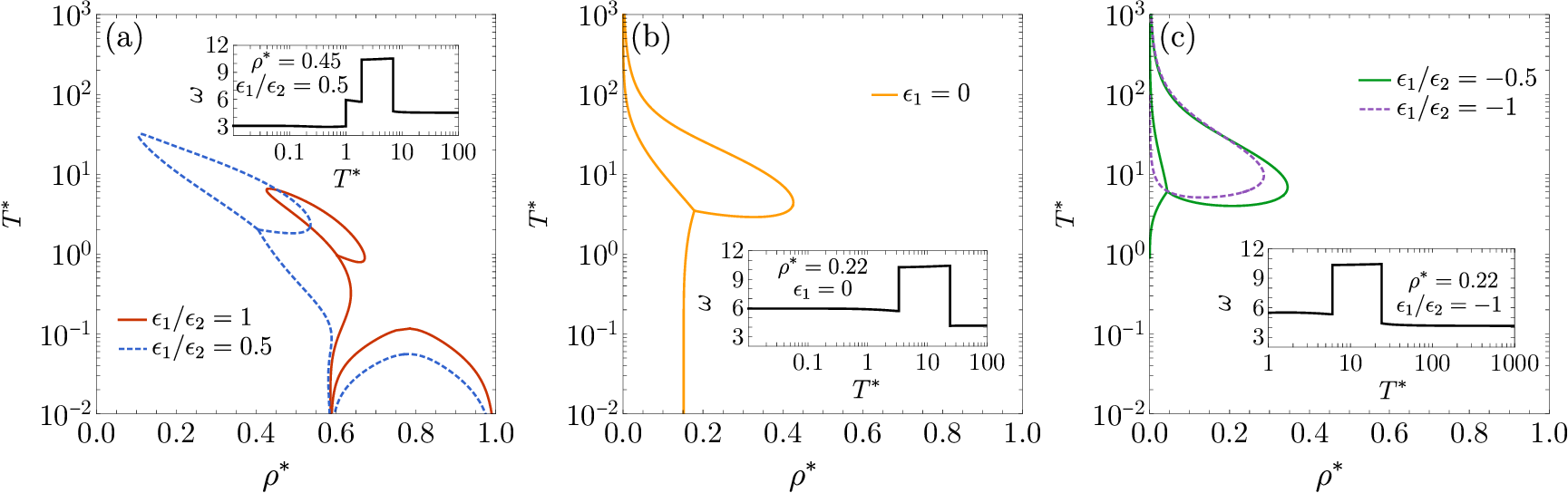}
\end{adjustwidth}

\vspace{-3pt}
\caption{
DOC lines on the $T^*$ vs. $\rho^*$ plane for the 1D case with $\lambda_1=1.35$ and $\lambda_2=1.7$: \mbox{(\textbf{a}) $\epsilon_1/\epsilon_2=1,0.5$}, (\textbf{b}) $\epsilon_1=0$, and (\textbf{c}) $\epsilon_1/\epsilon_2=-0.5,-1$. Insets display the angular frequency of the asymptotic oscillations of $h(r)$ as a function of $T^*$ for (\textbf{a}) $\epsilon_1/\epsilon_2=0.5$ at $\rho^*=0.45$, (\textbf{b}) $\epsilon_1=0$ at $\rho^*=0.22$, and (\textbf{c}) $\epsilon_1/\epsilon_2=-1$ at $\rho^*=0.22$.\label{fig3}}
\end{figure}

Figure \ref{fig3}{a} shows the DOC line, loop, and arc for $\epsilon_1 / \epsilon_2 = 1$ [also displayed in \mbox{Figure \ref{fig1}{b}}] and for $\epsilon_1 / \epsilon_2 = 0.5$. In the latter case, the loop expands and shifts up and to the left, while the arc moves downward.
For a fluid with $\epsilon_1 = 0$ [Figure \ref{fig3}{b}], the DOC line appears at a density below $\lambda_2^{-1}$, with the loop evolving into a lobe that emerges from the vertical axis at $\rho^* = 0$.
In Figure \ref{fig3}{c}, a short DOC line forms at very small densities when $\epsilon_1 / \epsilon_2 = -0.5$, but it vanishes when
$\epsilon_1 / \epsilon_2 = -1$. As $\epsilon_1 / \epsilon_2$ becomes increasingly negative, we have observed that the DOC lobe progressively contracts, moving up and to the
left until it eventually disappears. 
The insets in Figure \ref{fig3} show the oscillation frequency $\omega$ as a function of $T^*$ at selected values of density and the energy ratio $\epsilon_1 / \epsilon_2$.

Since the smallest length scale of the problem is the hard core diameter $\sigma=1$, one might reasonably expect the angular frequency of the asymptotic oscillations to remain below $\omega \approx 2\pi$. However, as discussed earlier, within the loops and arcs, $\omega$ is distinctly larger than $2\pi$, indicating wavelengths significantly shorter than the hard core diameter [see the insets in Figures \ref{fig1}{a},{b} and \ref{fig2}{b},{d}, as well as the insets in Figure \ref{fig3}]. This surprising phenomenon suggests the emergence of intricate, potentially novel mesoscopic ordering that warrants deeper investigation in future studies.

\subsection{$\lambda_1=1.35$, $\lambda_2=1.7$, and $\epsilon_1<0$: Influence of $\epsilon_1/\epsilon_2$ on  FW Line}\label{sec2.4}

We now consider the case $\epsilon_1 < 0$, where a genuine competition arises between the attractive square well with the depth $|\epsilon_1|$ and the repulsive barrier of the height $\epsilon_2$.  As demonstrated in Appendix \ref{appA1}, real poles of $G(s)$ may exist. If one of these real poles becomes dominant, the asymptotic decay of $h(r)$ is monotonic, and, as mentioned earlier, an FW line emerges, marking the abrupt transition between monotonic and oscillatory decay. However, a DOC line may still occur, as exemplified by Figure \ref{fig3}{c}.

\begin{figure}[H]

\includegraphics[width=0.8\columnwidth]{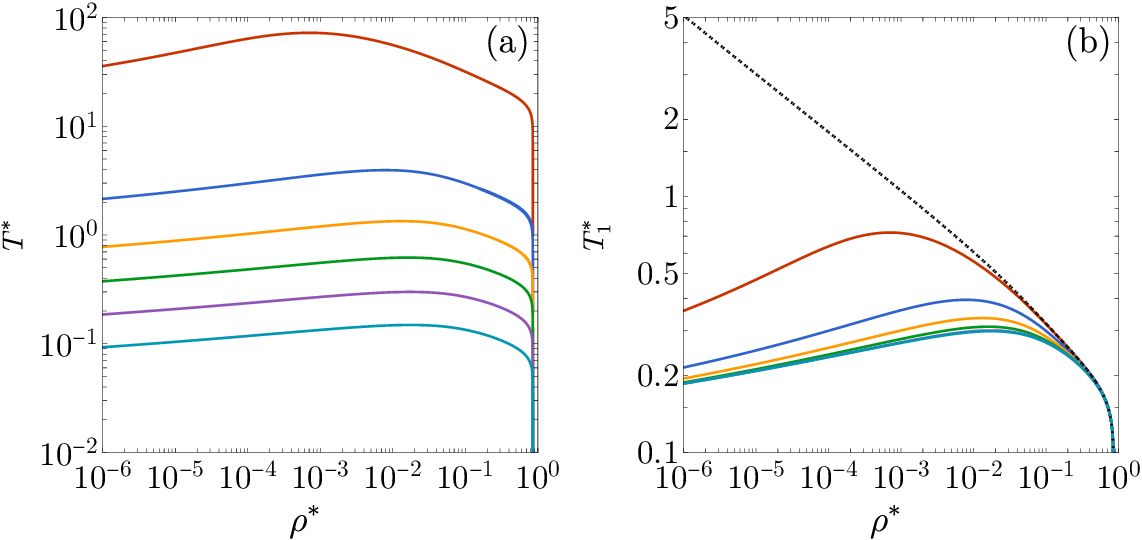}

\vspace{-3pt}
\caption{(\textbf{a}) {FW} 
 lines on the $T^*$ vs. $\rho^*$ plane for the 1D case $\lambda_1=1.35$ and $\lambda_2=1.7$ with, from the bottom to top, $\epsilon_1/\epsilon_2=-0.5$, $-1$, $-2$, $-4$, $-10$,  and $-100$.  (\textbf{b}) The same as in panel (\textbf{a}), except that now the vertical axis represents the scaled temperature $T_1^*= k_BT/|\epsilon_1| = T^*\epsilon_2/|\epsilon_1|$. The dotted curve is the FW line for a pure square-well fluid ($\epsilon_2=0$). Note that in panel (\textbf{b}) the curves corresponding to $\epsilon_1/\epsilon_2=-0.5$ and  $-1$ are indistinguishable.
\label{fig4}}
\end{figure}

The results for various values of $\epsilon_1/\epsilon_2<0$ are presented in Figure \ref{fig4}{a}. A comparison of the DOC lines in Figure \ref{fig3}{c} for $\epsilon_1/\epsilon_2 = -0.5$ and $-1$ with the corresponding FW lines in Figure \ref{fig4}{a} shows that the FW lines emerge at significantly lower values of $T^*$ and span a broader range of densities.

The strong sensitivity of the FW lines to the values of $\epsilon_1/\epsilon_2$, as seen in Figure \ref{fig4}{a}, is significantly reduced when the temperature is scaled by the well depth $|\epsilon_1|$, i.e., $T_1^* = k_B T / |\epsilon_1| = T^* \epsilon_2 / |\epsilon_1|$. This rescaling is applied in Figure \ref{fig4}{b}, which also includes the FW line for a pure square-well fluid ($\epsilon_2 / |\epsilon_1| \to 0$).
For the latter fluid, the FW line approaches $T_1^* \to \infty$ as $\rho^* \to 0$ (following a power law). However, introducing a repulsive barrier of the height $\epsilon_2$ causes the FW line to bend at low densities, even when $\epsilon_2 / |\epsilon_1| = 10^{-2}$.

It should be pointed out that in the 1D lattice model analyzed in Ref.~\cite{PCA13}, the attractive interaction is limited to nearest neighbors, while the repulsion extends up to third-nearest neighbors. At $T=0$, the energy minimum is achieved by forming clusters of three consecutive particles. The authors also report the formation of clusters separated by distances greater than the range of the repulsion. However, we observe neither of these features in the exact calculations of our model. Additionally, while an FW line is identified in their work, there is no evidence of a DOC.

\section{The 3D System: RFA Results}
\label{sec3}
\subsection{Theoretical Background}\label{sec3.1}

In this section, we provide a brief account of the main outcome of the RFA approach when the intermolecular potential in 3D is of the form of Equation (\ref{varphi}). The detailed derivation may be found in References \cite{SYH12,SYHBO13,YSH22}. We begin by considering a function, $G(s)$, which is distinct from its 1D counterpart. This function represents the Laplace transform of $r g(r)$; specifically,
\begin{equation}
\label{b3}
G(s)=\int_0^\infty {\dd} r\, {e}^{-rs} rg(r).
\end{equation}
We next define an
auxiliary function, $\Phi(s)$, directly related to $G(s)$ through
\beq
\label{b5}
G(s)=s\frac{\Phi(s)}{1+12\eta \Phi(s)},
\eeq
where $\eta=\frac{\pi}{6}\rho^*$ is the packing fraction.
Taking into account Equations \eqref{G(s)} and \eqref{Z}, we can say that $\Phi(s)$ is the 3D analog of the 1D quantity $\Omega(s+\bp)/\rho\Omega(\bp)$.
To reflect the discontinuities of $g(r)$ at the points $r=1$, $\la_1$, and $\la_2$, where $\varphi(r)$ is discontinuous, we decompose $\Phi(s)$ as
\beq
\Phi(s)=R_0(s)e^{-s}+R_1(s)e^{-\lambda_1 s}+R_2(s)e^{-\lambda_2 s}.
\label{3.1}
\eeq

Note that Equations \eqref{b3}--\eqref{3.1} are formally exact. Finally, to construct our RFA, we assume the following rational function approximate form for $R_j(s)$:
\beq
R_j(s)=-\frac{1}{12\eta}\frac{A_j+B_j s}{1+S_1 s+S_2 s^2+S_3 s^3}, \quad j=0,1,2.
\label{c6}
\eeq
The approximation in \eqref{c6} contains nine parameters to be determined by the application of certain constraints \cite{SYH12}. The expressions for those nine coefficients are presented in Appendix \ref{appB}.

Once again, the total correlation function $h(r)$ can be expressed in terms of the nonzero poles of $G(s)$, which, in principle, form an infinite set. These poles may be either real or occur in complex--conjugate pairs. Their locations depend on the thermodynamic state, and as before, the pole with the real part closest to zero dictates the asymptotic behavior of the total correlation function for a given state.
The 3D analogs of Equations \eqref{h(r)_1D_1} and \eqref{h(r)_1D} are, respectively,
\begin{subequations}
\begin{equation}
\label{h(r)_3D_1}
h(r)\approx\frac{1}{r}\left[
2|A_{\zeta_1}|e^{-\zeta_1 r}\cos(\omega_1 r+\delta_1)+
2|A_{\zeta_2}|e^{-\zeta_2 r}\cos(\omega_2 r+\delta_2)\right],  \quad r\gg 1
\end{equation}
\begin{equation}
\label{h(r)_3D}
h(r)\approx\frac{1}{r}\left[
2|A_\zeta|e^{-\zeta r}\cos(\omega r+\delta)+
A_\kappa e^{-\kappa r}\right], \quad r\gg 1.
\end{equation}
\end{subequations}
In the context of the RFA, Equations \eqref{b5}--\eqref{c6} imply that the complex poles satisfy the following set of coupled equations:
\vspace{-9pt}
\begin{adjustwidth}{-\extralength}{0cm}
 \centering 
\begin{subequations}
\beq
1-(S_1-S_2\zeta+S_3\zeta^2)\zeta-(S_2-3S_3\zeta)\omega^2=\sum_{j=0}^2e^{\zeta\lambda_j}\left[\left(A_j-B_j\zeta\right)\cos(\omega\lambda_j)+B_j\omega\sin(\omega\lambda_j)\right],
\eeq
\beq
-(S_1-2S_2\zeta+3S_3\zeta^2)\omega+S_3\omega^3=\sum_{j=0}^2e^{\zeta\lambda_j}\left[\left(A_j-B_j\zeta\right)\sin(\omega\lambda_j)-B_j\omega\cos(\omega\lambda_j)\right],
\eeq
\end{subequations}
\end{adjustwidth}
with the convention $ \lambda_0=1$.
Analogously, the real poles are the roots of
\beq
1-(S_1-S_2\kappa+S_3\kappa^2)\kappa=\sum_{j=0}^2e^{\kappa\lambda_j}\left(A_j-B_j\kappa\right).
\eeq

It should be noted that the RFA results become less reliable at lower temperatures and/or higher densities. Therefore, we will primarily focus on cases where $T^* > 0.5$ and $\rho^*<0.6$.
We now present our results following the same structure as in the 1D case (see Section \ref{sec2}).

\subsection{$\epsilon_1=\epsilon_2>0$: Influence of $\lambda_2$ on DOC Line}\label{sec3.2}

\begin{figure}[H]

\includegraphics[width=0.8\columnwidth]{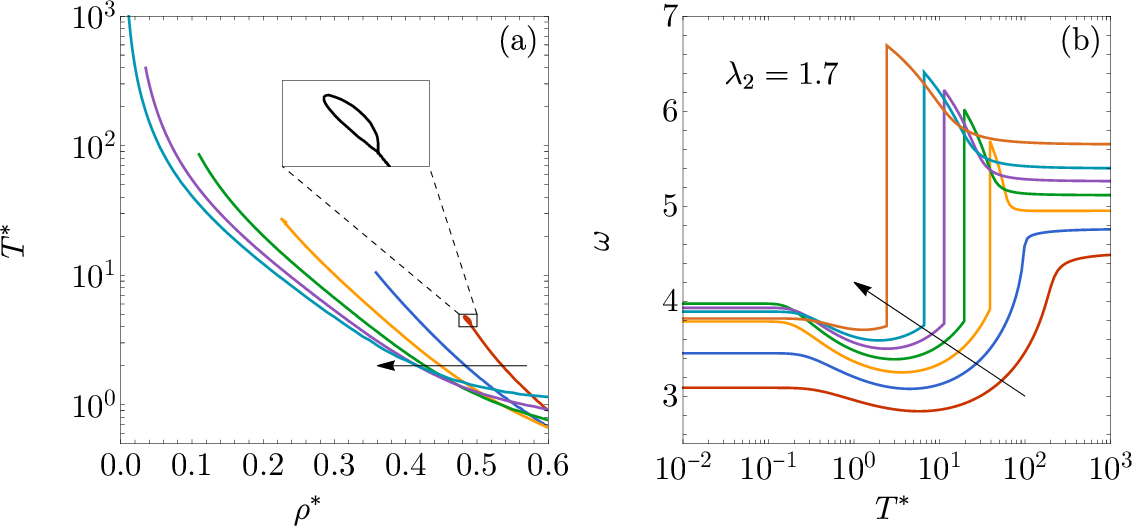}
\caption{(\textbf{a}) {DOC} 
 lines on the $T^*$ vs. $\rho^*$ plane for the 3D case $\epsilon_1=\epsilon_2$ with $\lambda_2=1.55$, $1.6$, $1.65$, $1.7$, $1.75$, and $1.8$.  The inset shows the loop corresponding to $\lambda_2=1.55$. (\textbf{b}) The angular frequency of the asymptotic oscillations of $h(r)$ plotted as a function of $T^*$ for $\rho^*=0.05$, $0.1$, $0.15$, $0.2$, $0.25$, $0.3$, and $0.4$, with an interaction potential characterized by  $\epsilon_1=\epsilon_2$ and $\lambda_2=1.7$. The arrows indicate the direction of increasing (\textbf{a}) $\lambda_2$ and (\textbf{b}) $\rho^*$.
\label{fig5}}
\end{figure}

Figure \ref{fig5}{a} displays the DOC lines for 3D fluids with $\epsilon_1 = \epsilon_2 > 0$, corresponding to values of $\lambda_2 = 1.55$, $1.6$, $1.65$, $1.7$, $1.75$, and $1.8$. The overall shape of these lines is qualitatively similar to the lines shown in Figure \ref{fig1}{b} for $\lambda_2 = 1.8$ and $1.7$ but has noticeably smaller loops, particularly as $\lambda_2$ increases. Within these loops, as in the 1D case, the oscillation frequency  is approximately $\omega/2\pi \approx 3/\lambda_2$. Furthermore, the DOC arcs observed in Figure \ref{fig1}{b},{c} for 1D fluids are absent in Figure \ref{fig5}{a}, as they would be confined to the high-density, low-temperature region where the RFA is no longer reliable. Indeed, no DOC line is observed for $\lambda_2 \leq 1.5$, consistent with the disappearance of the single DOC line in Figure \ref{fig1}{c} for $\lambda_2 = 1.6$ and $1.5$. Additionally, the 3D density playing the role of the 1D value $\rho^* = \lambda_2^{-1}$ is given by $\rho^* = \rho^*_{\max} \lambda_2^{-3}$, where $\rho^*_{\max} \simeq 0.94$ represents the freezing density of hard spheres \cite{AW57}.

A comparison between Figure \ref{fig5}{b} and  the inset of Figure \ref{fig1}{a} reveals a shared characteristic: when the single DOC line is crossed at a given density while moving from higher to lower temperatures, the frequency $\omega$ initially increases near the crossover temperature before suddenly dropping to a smaller value.

\subsection{$\lambda_1=1.35$ and $\lambda_2=1.7$: Influence of $\epsilon_1/\epsilon_2$ on DOC Line}\label{sec3.3}

\begin{figure}[H]

\includegraphics[width=0.8\columnwidth]{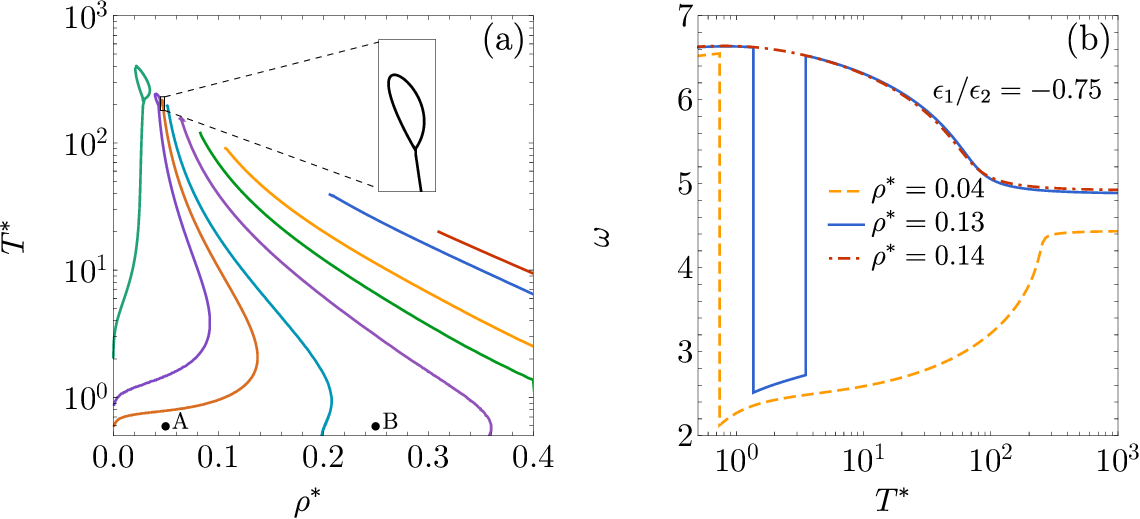}
\caption{(\textbf{a}) {DOC} 
 lines on the $T^*$ vs. $\rho^*$ plane for the 3D case $\lambda_1=1.35$ and $\lambda_2=1.7$ with, from right to left, $\epsilon_1/\epsilon_2=2.5$, $2$, $1$, $0.5$, $0$, $-0.5$, $-0.75$,  $-1$, and $-2$. The inset shows the loop corresponding to $\epsilon_1/\epsilon_2=-0.75$. (\textbf{b}) The angular frequency of the asymptotic oscillations of $h(r)$ plotted as a function of $T^*$ for $\rho^*=0.04$, $0.13$,  and $0.14$, with an interaction potential characterized by  $\epsilon_1/\epsilon_2=-0.75$, $\lambda_1=1.35$, and $\lambda_2=1.7$. The circles in panel (\textbf{a}) represent the two states examined in Figure \ref{fig6plus} for $\epsilon_1/\epsilon_2=-0.5$.
\label{fig6}}
\end{figure}

Figure \ref{fig6}{a} displays the DOC lines on the $T^*$ vs. $\rho^*$ plane for various values of $\epsilon_1/\epsilon_2$, covering the cases where $\epsilon_1>0$, $\epsilon_1=0$, and $\epsilon_1<0$.
In analogy with the 1D case [see Figure \ref{fig3}{b},{c}], these lines exhibit qualitative changes as the system transitions from positive to negative values of $\epsilon_1$.
However, in the 3D case, the loops apparently do not degenerate into lobes.

Another notable feature is the rounded, bulging profile of the DOC lines for \mbox{$\epsilon_1 < 0$}. This shape indicates that, within a certain density interval, the frequency $\omega$ exhibits reentrant behavior as the temperature varies. This phenomenon is illustrated in \mbox{Figure \ref{fig6}{b}} for $\epsilon_1/\epsilon_2 = -0.75$.
At the density $\rho^* = 0.04$ (below the loop densities $\rho^* \approx 0.05$), the oscillation frequency undergoes a single drop from $\omega \simeq 6.6$ to $\omega \simeq 2.1$ when crossing the temperature $T^* \simeq 0.76$ from left to right and then increases smoothly toward $\omega \simeq 4.4$ at the high-temperature limit.
At a higher density, $\rho^* = 0.13$ (just below the bulge's end at $\rho^* \simeq 0.138$), a more complex behavior is observed: the frequency drops from $\omega \simeq 6.6$ to $\omega \simeq 2.5$ at $T^* \simeq 1.3$ and then rises again from $\omega \simeq 2.7$ to $\omega \simeq 6.5$ at $T^* \simeq 3.6$, eventually tending smoothly toward $\omega \simeq 4.9$ at high temperatures.
Finally, at $\rho^* = 0.14$, the evolution of $\omega$ from $\omega \simeq 6.6$ at low temperatures to $\omega \simeq 4.9$ at high temperatures proceeds continuously without reentrant behavior.

\begin{figure}[H]

\includegraphics[width=0.8\columnwidth]{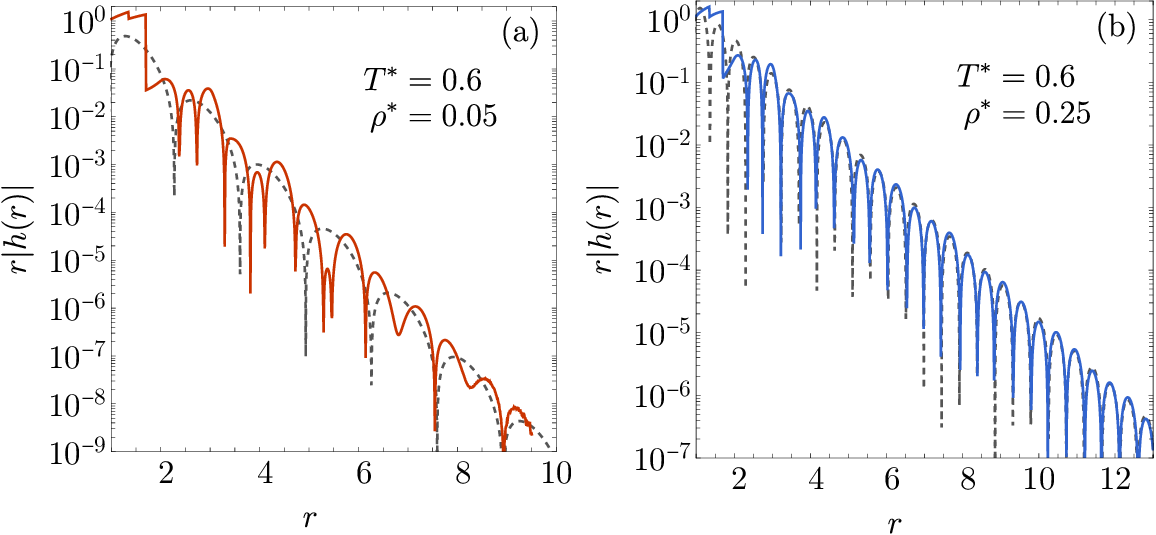}
\caption{A logarithmic plot of $r |h(r)|$  in the 3D case $\epsilon_1/\epsilon_2=-0.5$, $\lambda_1=1.35$, and $\lambda_2=1.7$,  for the following states: (\textbf{a}) $(\rho^*,T^*)=(0.05,0.6)$ and (\textbf{b}) $(\rho^*,T^*)=(0.25,0.6)$. These states are labeled A and B in Figure \ref{fig6}{a}, respectively.
The solid lines illustrate the values derived from numerical Laplace inversion, whereas the dashed lines depict the asymptotic expression $r |h(r)| = 2 e^{-\zeta r}|A_\zeta\cos(\omega r + \delta)|$, where $s= -\zeta \pm \imath \omega$ denotes the leading pole of $G(s)$.\label{fig6plus}}
\end{figure}

In a manner analogous to the 1D case (see Figure \ref{fig2}), we numerically invert the Laplace transform defined by Equations \eqref{b5}--\eqref{c6} using the method outlined in \cite{EulerILT} to derive $h(r)$. The results for $\epsilon_1/\epsilon_2 = -0.5$, $\lambda_1 = 1.35$, and $\lambda_2 = 1.7$ are presented in Figure \ref{fig6plus} for two representative states, labeled A and B in Figure \ref{fig6}{a}. Additionally, \mbox{Figure \ref{fig6plus}} includes the asymptotic form $r |h(r)| = 2 e^{-\zeta r}|A_\zeta\cos(\omega r + \delta)|$ with the parameters \mbox{$(\zeta, \omega) = (2.324, 2.364)$} for state A and $(\zeta, \omega) = (1.283, 6.714)$ for state B. The competition between the leading and subleading poles is evident in Figure \ref{fig6plus}{a}, where the leading-pole asymptotic behavior requires distances greater than $r \approx 10$. In contrast, for state B, the asymptotic behavior is effectively reached beyond $r \approx 3$. Overall, the contrast between low- and high-frequency oscillations is clearly observed to the left and right of the DOC line, respectively.

The phenomenon of $\omega$ being distinctly larger than $2\pi$ within the loops persists in 3D fluids. However, we have verified that, because these loops are much smaller in size compared to the 1D case, the competition between the leading and subleading poles causes the asymptotic one-pole behavior to dominate only at very large distances. At such scales, the amplitude of the oscillations of $|h(r)|$ can diminish to extremely small values, potentially below $10^{-10}$.

\subsection{$\lambda_1=1.35$, $\lambda_2=1.7$, and $\epsilon_1<0$: Influence of $\epsilon_1/\epsilon_2$ on FW Line}\label{sec3.4}

\begin{figure}[H]

\includegraphics[width=0.8\columnwidth]{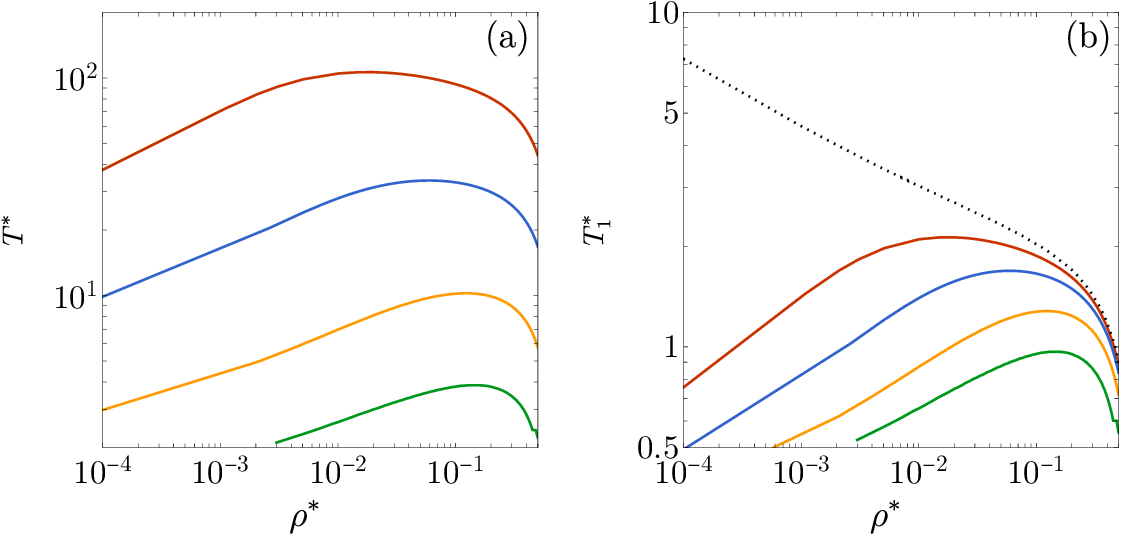}
\caption{(\textbf{a}) {FW} 
 lines on the $T^*$ vs. $\rho^*$ plane for the 3D case $\lambda_1=1.35$ and $\lambda_2=1.7$ with, from the bottom to top,  $\epsilon_1/\epsilon_2=-4$, $-8$, $-20$,   and $-50$.   (\textbf{b}) The same as in panel (\textbf{a}), except that now the vertical axis represents the scaled temperature $T_1^*= k_BT/|\epsilon_1| = T^*\epsilon_2/|\epsilon_1|$. The dotted line is the FW line for a pure square-well fluid ($\epsilon_2/|\epsilon_1|=0$).
\label{fig7}}
\end{figure}

In the 1D case, an FW line is already observed with $\epsilon_1/\epsilon_2 = -0.5$, but this requires temperatures in the order of $T^* \sim 10^{-1}$ [see Figure \ref{fig4}{a}]. Since, as mentioned earlier, the RFA tends to provide less reliable results at low temperatures, it becomes necessary to consider deeper wells to study the FW lines for 3D fluids.
The cases $\epsilon_1/\epsilon_2 = -4$, $-8$, $-20$, and $-50$ are reported in Figure \ref{fig7}{a}. As in the 1D fluid, it is useful to plot the curves on the $T_1^*$ vs. $\rho^*$ plane to compare them with the FW line of the pure square-well fluid, as shown in Figure \ref{fig7}{b}. Again, we observe that the presence of the repulsive barrier between $\lambda_1$ and $\lambda_2$ bends the FW line downward for the low-density region. This indicates that the decay of $h(r)$ is always oscillatory when the temperature exceeds a certain threshold, regardless of the density.

\section{Conclusions}
\label{sec4}
In this paper, we explore the impact of competing interactions in the intermolecular potential of fluids on their structural transitions. The model potential adopted for both 1D and 3D systems consists of a hard core followed by two steps, which can represent either a shoulder or a well depending on the sign of the parameters $\epsilon_1$ and $\epsilon_2$. This potential is versatile enough to encompass a range of competing interactions, including the square-well and square-shoulder interactions as limiting cases. Additionally, the consideration of two different dimensionalities allows us to examine the influence of strong confinement on the structural transitions of these fluids. For the 1D systems, restricting the interaction range to no more than twice the hard core diameter enables us to derive exact results. In contrast, for the 3D systems, where exact solutions are not feasible, we employ the RFA to obtain and analyze approximate structural properties.

The results for both the 1D and 3D systems align with the expected behavior. Specifically, at very low temperatures, the decay of the total correlation function $h(r)$ exhibits oscillations with a wavelength determined by the range of the repulsive barrier, provided that both $\epsilon_1$ and $\epsilon_2$ are positive. In contrast, at very high temperatures, the oscillations have a wavelength on the order of the hard core diameter. Furthermore, it is confirmed that at a given density, the transition between these two regimes as the temperature increases can occur either continuously or, as observed in binary hard-sphere mixtures, discontinuously upon crossing a DOC line.

When $\epsilon_1$ is negative, an FW transition from an oscillatory to a monotonic decay of $h(r)$ occurs as the temperature decreases at a given density, even when $\epsilon_2$ is positive. Additionally, the presence of the repulsive barrier of the height $\epsilon_2$ causes the FW line to exhibit a maximum at a certain density before bending downward at lower densities, in stark contrast to its behavior in the absence of such a barrier.

While the results for both the 1D and 3D systems exhibit many common characteristic features, the effects of dimensionality introduce notable distinctions. These include shifts in the temperature ranges in which certain features appear, the need for deeper wells to observe similar phenomena, or a reduction in their prominence as the system transitions from 1D to 3D. In some cases, features present in 1D may vanish entirely in 3D. Notably, we emphasize the complex behavior of the DOC transition, as previously discussed. This intricacy manifests in phenomena such as loops, arcs, lobes, triple points, and reentrant frequencies, some of which, to the best of our knowledge, have not been reported in this context before.

{In summary, we uncovered a remarkably complex pattern of structural transitions in fluids with intermolecular potentials that include competing interactions. Even for the relatively simple potential considered in this work, analyzing structural transitions required exploring a broad (dimensionless) parameter space, involving $\lambda_1$, $\lambda_2$, $\epsilon_1/\epsilon_2$, $T^*$, and $\rho^*$.
Given these circumstances, our findings are undoubtedly limited. Nevertheless, they reveal a fascinating and intricate phenomenology that merits further and more detailed exploration. In particular, establishing a connection between this phenomenology and the structures and patterns observed in SALR fluids remains an open and compelling~challenge.}

\vspace{6pt}
\authorcontributions{Conceptualization, A.S.; methodology, S.B.Y., A.S. and M.L.H.; software, A.M.M. and S.B.Y.; validation, A.M.M., S.B.Y., A.S. and M.L.H.; formal analysis, A.S. and M.L.H.; investigation, A.M.M., S.B.Y., A.S. and M.L.H.;  writing---original draft preparation, A.S. and M.L.H.; writing---review and editing, A.M.M., S.B.Y., A.S. and M.L.H.; visualization, A.M.M.; supervision, S.B.Y., A.S. and M.L.H.; funding acquisition, S.B.Y. and A.S. All authors have read and agreed to the published version of the manuscript.}

\funding{This research was funded by  MCIN/AEI/10.13039/501100011033, grant number PID2020-112936GB-I00. A.M.M. is grateful to the Spanish Ministerio de Ciencia e Innovaci\'on for a predoctoral fellowship, grant no.\ PRE2021-097702.}

\institutionalreview{{~}Not applicable. 
}

\dataavailability{The raw data supporting the conclusions of this article will be made available by the authors on request.}

\conflictsofinterest{The authors declare no conflicts of interest.}

\abbreviations{Abbreviations}{
The following abbreviations are used in this manuscript:\\

\noindent
\begin{tabular}{@{}ll}
1D& One-dimensional\\
3D& Three-dimensional\\
DOC& Discontinuous oscillation crossover\\
FW & Fisher--Widom\\
HR & Hard rod\\
RFA & Rational function approximation\\
SALR & Short-range attraction and long-range repulsion
\end{tabular}
}

\appendixtitles{yes} 
\appendixstart
\appendix

\section{Some Mathematical Details in the Case of the 1D Fluid }
\label{appA}

\subsection{Absence of Real Poles If $\varphi(r)\geq 0$}
\label{appA1}
We first consider a generic potential, $\varphi(r)$, that goes to $\infty$ if $r<\sigma$ and vanishes if $r>b>\sigma$. If a real pole, $s=-\kappa< 0$, of $G(s)$ exists, then
\beq
\label{condi}
\Omega(\bp-\kappa)-\Omega(\bp)=0.
\eeq

If $s$ is real and positive, $\Omega(s)$ decreases monotonically with $s$. Consequently, \mbox{Equation~\eqref{condi}} cannot be satisfied for $0 < \kappa \leq \bp$.
If, on the other hand, $\kappa >\bp$, the argument $s=\bp-\kappa$ becomes negative and, thus, we first need to evaluate $\Omega(s)$ assuming that $s>0$ and then perform an analytic continuation to $s<0$. With $s>0$,
\beq
\Omega(s)=\int_0^b \dd r\, e^{-s r}\left[e^{-\beta\varphi(r)}-1\right]+\frac{1}{s}.
\eeq
This expression can now be analytically continued to $s<0$. Therefore, if $\kappa>\bp$, we \mbox{can write}
\beq
\Omega(\bp-\kappa)-\Omega(\bp)=\int_0^b \dd r\, e^{-\bp r}\left(e^{\kappa r}-1\right)\left[e^{-\beta\varphi(r)}-1\right]-\left(\frac{1 }{\kappa-\bp}+\frac{1}{\bp}\right).
\eeq
If $\varphi(r)\geq 0$, then $e^{-\beta\varphi(r)}-1\leq 0$. In that case, and given that $\kappa>\bp$, one has $\Omega(\bp-\kappa)-\Omega(\bp)<0$ and Equation \eqref{condi} cannot be satisfied.

In summary, if $\varphi(r)\geq 0$ for all $r$, then no real poles exist. Otherwise,  the real poles $s=-\kappa$ may exist and, then, $\kappa>\bp$.

\subsection{Poles for the High-Temperature Limit}
\label{appA2}
For the limit $\beta \to 0$, the system simplifies to an HR fluid  with the diameter $\sigma = 1$. In this regime, Equations \eqref{EOS} and \eqref{eq:poles} reduce to
\begingroup\makeatletter\def\f@size{9}\check@mathfonts
\def\maketag@@@#1{\hbox{\m@th\normalsize\normalfont#1}}
\begin{subequations}
\beq
\bp=\frac{\rho^*}{1-\rho^*},
\eeq
\beq
\label{Real}
1-\frac{1-\rho^*}{\rho^*}\zeta_\hr(\rho^*)=e^{\zeta_\hr(\rho^*)}\cos\omega_\hr(\rho^*),\quad
-\frac{1-\rho^*}{\rho^*}\omega_\hr(\rho^*)= e^{\zeta_\hr}(\rho^*)\sin\omega_\hr(\rho^*).
\eeq
\end{subequations}\endgroup
By squaring both sides of the equalities in Equation \eqref{Real} and adding them, $\omega_\hr$ can be expressed as a function of $\zeta_\hr$:
\beq
\label{omega}
\omega_\hr(\rho^*)=\frac{\rho^*}{1-\rho^*}\sqrt{e^{2\zeta_\hr(\rho^*)}-\left[1-\frac{1-\rho^*}{\rho^*}\zeta_\hr(\rho^*)\right]^2}.
\eeq
Substituting this expression into the first equality of Equation \eqref{Real} yields a closed equation for $\zeta_\hr$, which can be solved numerically to find the pole with the smallest value of $\zeta_\hr$. It is important to discard any spurious root that may appear for $\rho^* < 1/2$ as a consequence of squaring the equations.

\begin{figure}[H]

\includegraphics[width=0.8\columnwidth]{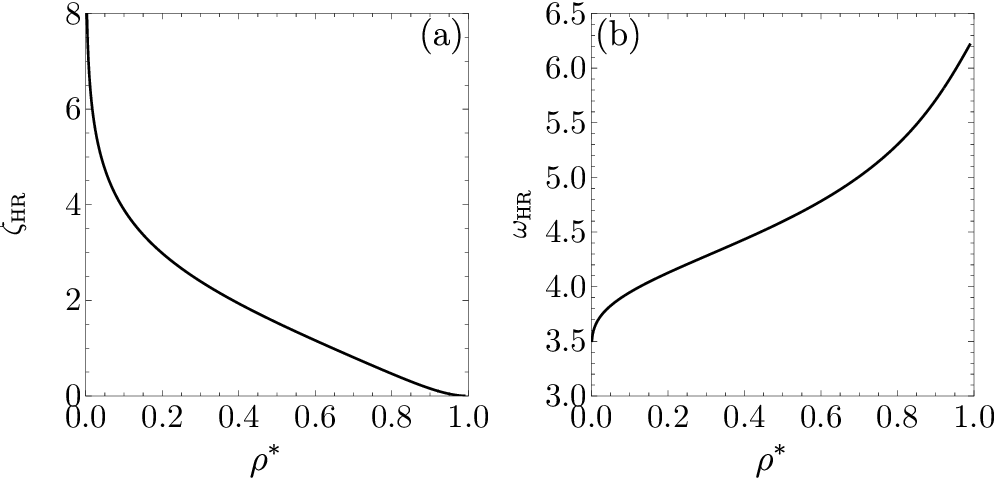}
\caption{Density dependence of (\textbf{a}) $\zeta_\hr(\rho^*)$  and (\textbf{b}) $\omega_\hr(\rho^*)$.
\label{figA1}}
\end{figure}

For the close-packing limit $\rho^*\to 1$, it is easy to obtain
\beq
\omega_\hr(\rho^*)\approx 2\pi \rho^*,\quad \zeta_\hr(\rho^*)\approx 2\pi^2(1-\rho^*)^2.
\eeq
On the other hand, for the opposite low-density limit ($\rho^*\to 0$),  one has $\omega_\hr\approx \pi$ and  $\zeta_\hr e^{-\zeta_\hr}=\rho^*$. The latter belongs to the class of Lambert equations $z=w e^w$ with $z=-\rho^*$ and $w=-\zeta_\hr$. The solution is then
\beq
\zeta_\hr(\rho^*)\approx -W_{-1}(-\rho^*),\quad
\omega_\hr(\rho^*)\approx \pi\left[1+\zeta_\hr^{-1}(\rho^*)\right],
\eeq
where $W_{-1}(z)$ is the lower branch of the Lambert function \cite{CGHJK96}. The functions $\zeta_\hr(\rho^*)$  and  $\omega_\hr(\rho^*)$ are plotted in Figure \ref{figA1}.

\subsection{Poles for the Low-Temperature Limit with $\epsilon_1>0$}
\label{appA3}
\subsubsection{Case $0<\rho^*<\lambda_2^{-1}$}
\label{appA3.1}
For the low-temperature limit $\beta \to \infty$ with $\epsilon_1>0$, one can see from \mbox{Equations \eqref{Omega}--\eqref{eq:poles}} that the system becomes equivalent to an HR fluid with the diameter $\lambda_2$, provided that \mbox{$\rho^*<\lambda_2^{-1}$}. Therefore, for that limit, one has
\beq
\omega(\rho^*)=\lambda_2^{-1}\omega_\hr(\rho^*\lambda_2),\quad \zeta(\rho^*)=\lambda_2^{-1}\zeta_\hr(\rho^*\lambda_2).
\eeq

\subsubsection{Case $\lambda_2^{-1}<\rho^*<1$}
\label{appA3.2}

For the limit  $\beta\to\infty$ with finite $p$, one has $e^{-\bp}\gg e^{-\bp\lambda_1}\gg e^{-\bp\lambda_2}$, so Equation \eqref{EOS} yields
\beq
\label{13b}
\rho^*=\frac{e^{-\beta(\epsilon_1+p)}+e^{-\beta(\epsilon_2+p\lambda_1)}+e^{- \bp\lambda_2}}{e^{-\beta(\epsilon_1+p)}+\lambda_1e^{-\beta(\epsilon_2+p\lambda_1)}+\lambda_2e^{- \bp\lambda_2}}.
\eeq
The dominant terms in Equation \eqref{13b} depend on the domain of $p$.
One can distinguish three possibilities.
First, if
\beq
\label{A6}
p\lambda_2<\min\{\epsilon_1+p,\epsilon_2+p\lambda_1\},
\eeq
one has  $\rho^*=\lambda_2^{-1}$.
Next, if
\beq
\label{A9}
\epsilon_1+p<\min\{p\lambda_2,\epsilon_2+p\lambda_1\},
\eeq
the result is $\rho^*=1$.
Finally, if
\beq
\label{A11}
\epsilon_2+p\lambda_1<\min\{p\lambda_2,\epsilon_1+p\},
\eeq
then
 $\rho^*=\lambda_1^{-1}$.

Suppose first that $0<\epsilon_1/\epsilon_2\leq (\lambda_2-1)/(\lambda_2-\lambda_1)$. In that case, the inequalities in \eqref{A6} and \eqref{A9} imply $p<\epsilon_1/(\lambda_2-1)$ and $p>\epsilon_1/(\lambda_2-1)$, respectively, while the inequality in \eqref{A11} is impossible.
Thus, for the low-temperature limit, the density changes from $\rho^*=\lambda_2^{-1}$ to $\rho^*=1$ as the pressure crosses the value $p=\epsilon_1/(\lambda_2-1)$. To analyze this change in detail, let us introduce the scaled variable $\mu$ by
\beq
\label{A14}
p=\frac{\epsilon_1}{\lambda_2-1}\left(1+\frac{\mu}{\beta\epsilon_1}\right),
\eeq
so that Equation \eqref{13b} becomes
\beq
\rho^*=\frac{1+e^{\mu}}{\lambda_2+e^{\mu}},\quad \mu=\ln\frac{\rho^*\lambda_2-1}{1-\rho^*}.
\eeq
In turn, from Equation \eqref{eq:poles}, we have
\begin{subequations}
\label{A16}
\beq
1-\frac{\zeta}{\beta\epsilon_1}(\lambda_2-1)=e^\zeta\frac{(\rho^*\lambda_2-1)\cos\omega+(1-\rho^*)e^{\zeta(\lambda_2-1)}\cos(\omega\lambda_2)}{\rho^*(\lambda_2-1)},
\eeq
\beq
-\frac{\omega}{\beta\epsilon_1}(\lambda_2-1)=e^\zeta\frac{(\rho^*\lambda_2-1)\sin\omega+(1-\rho^*)e^{\zeta(\lambda_2-1)}\sin(\omega\lambda_2)}{\rho^*(\lambda_2-1)}.
\eeq
\end{subequations}
For simplicity, let us assume that $\lambda_2$ is a rational number, $\lambda_2=m_2/n_2$. The analytical solution to Equations \eqref{A16} in the limit $\beta\epsilon_1\to\infty$ is displayed in the first row of Table \ref{tab1}.

\begin{table}[H]
\footnotesize
\caption{Asymptotic expressions for $p$, $\zeta$, and $\omega$ for the low-temperature limit for 1D fluids with $\epsilon_j>0$, assuming $\lambda_1 = m_1 / n_1$, $\lambda_2 = m_2 / n_2$, and $\lambda_2 / \lambda_1 = m_{21} / n_{21}$ are rational numbers.\label{tab1}}
\newcolumntype{C}{>{\centering\arraybackslash}X}
\begin{adjustwidth}{-\extralength}{0cm}
\newcolumntype{C}{>{\centering\arraybackslash}X}
\begin{tabularx}{\fulllength}{m{2cm}<{\centering}m{2.5cm}<{\centering}ccc}
\toprule
\boldmath{$\epsilon_1/\epsilon_2$}	& \boldmath{$\rho^*$}	& \boldmath{$p$} & \boldmath{$\zeta$} & \boldmath{$\omega$}\\
\midrule
$\displaystyle{\frac{\epsilon_1}{\epsilon_2}<\frac{\lambda_2-1}{\lambda_2-\lambda_1}}$		& $\lambda_2^{-1}<\rho^*<1$	 &
$\displaystyle{\frac{\epsilon_1}{\lambda_2-1}\left(1+\frac{\ln\frac{\rho^*\lambda_2-1}{1-\rho^*}}{\beta\epsilon_1}\right)}$ &
$\displaystyle{\frac{2[n_2\pi(\lambda_2-1)\rho^*]^2}{(\beta \epsilon_1)^2}\left[1+\lambda_2(1-\rho^*)\right]}$   &$\displaystyle{2n_2\pi\left[1-\frac{(\lambda_2-1)\rho^*}{\beta\epsilon_1}\right]}$\\
$\displaystyle{\frac{\epsilon_1}{\epsilon_2}>\frac{\lambda_2-1}{\lambda_2-\lambda_1}}$		& $\lambda_2^{-1}<\rho^*<\lambda_1^{-1}$			& $\displaystyle{\frac{\epsilon_2}{\lambda_2-\lambda_1}\left(1+\frac{\ln\frac{\rho^*\lambda_2-1}{1-\rho^*\lambda_1}}{\beta\epsilon_2}\right)}$ &
$\displaystyle{\frac{2[n_{21}\pi(\lambda_2-\lambda_1)\rho^{*}]^2}{(\beta \epsilon_2\lambda_1)^2}\left[\lambda_1+\lambda_2(1-\rho^*\lambda_1)\right]}$&
$\displaystyle{\frac{2n_{21}\pi}{\lambda_1}\left[1-\frac{(\lambda_2-\lambda_1)\rho^*}{\beta\epsilon_2}\right]}$ \\
& $\lambda_1^{-1}<\rho^*<1$	 & $\displaystyle{\frac{\epsilon_1-\epsilon_2}{\lambda_1-1}\left[1+\frac{\ln\frac{\rho^*\lambda_1-1}{1-\rho^*}}{\beta(\epsilon_1-\epsilon_2)}\right]}$&
$\displaystyle{\frac{2[n_1\pi(\lambda_1-1)\rho^*]^2}{[\beta (\epsilon_1-\epsilon_2)]^2}\left[1+\lambda_1(1-\rho^*)\right]}$ &$\displaystyle{2n_1\pi\left[1-\frac{(\lambda_1-1)\rho^*}{\beta(\epsilon_1-\epsilon_2)}\right]}$\\
\bottomrule
\end{tabularx}
\end{adjustwidth}
\end{table}

In contrast, for a potential where $\epsilon_1/\epsilon_2 > (\lambda_2 - 1)/(\lambda_2 - \lambda_1)$, the situation becomes more intricate. In this case, the inequalities in Equations \eqref{A6}--\eqref{A11} imply the following conditions: $ p < \epsilon_2 / (\lambda_2 - \lambda_1)$; $ p > (\epsilon_1 - \epsilon_2) / (\lambda_1 - 1) $; and $ \epsilon_2 / (\lambda_2 - \lambda_1) < p < (\epsilon_1 - \epsilon_2) / (\lambda_1 - 1) $, respectively. This indicates that, for the low-temperature limit, the density changes from $ \rho^* = \lambda_2^{-1}$ to $\rho^* = \lambda_1^{-1}$ as the pressure crosses $ p = \epsilon_2 / (\lambda_2 - \lambda_1)$ and from $\rho^* = \lambda_1^{-1}$ to $\rho^* = 1$ as the pressure crosses $p = (\epsilon_1 - \epsilon_2) / (\lambda_1 - 1)$. An analysis similar to that conducted in Equations \eqref{A14}--\eqref{A16} leads to the expressions shown in the second and third rows of Table \ref{tab1}.
An oscillation discontinuity at $\rho = \lambda_1^{-1}$ only arises if $n_{21} \neq m_1$, meaning that $n_1 / n_2$ must be an integer. This condition excludes cases such as $\lambda_1 = 1.35$ and $\lambda_2 = 1.7$, where $n_1 = 20$, $m_1 = 27$, $n_2 = 10$, $m_2 = 17$, $n_{21} = 27$, and $m_{21} = 34$.

\section{Parameters in Equation \eqref{c6}}\label{appB}
First, the exact condition $G(s)=s^{-2}+\mathcal{O}(s^0)$ for small $s$ yields
\begin{subequations}
\label{c9}
\beq
1=A_0+ A_1+A_2,\quad
S_1=-1+B_0-C^\one,\quad
S_2=\frac{1}{2}-B_0+C^\one+\frac{1}{2}C^\two,
\eeq
\beq
S_3=-\frac{1+2\e}{12\e}+\frac{1}{2}B_0-\frac{1}{2}C^\one-\frac{1}{2}C^\two-\frac{1}{6}C^\three,
\label{c10}
\eeq
\beq
B_0=C^\one+\frac{\e/2}{1+2\e}\left[6C^\two+4C^\three+C^\four\right]+\frac{1+\e/2}{1+2\e}.
\label{c11}
\eeq
\end{subequations}
Here,
\beq
C^{(k)}\equiv A_1 (\lambda_1-1)^k+A_2 (\lambda_2-1)^k -k B_1(\lambda_1-1)^{k-1}-k B_2(\lambda_2-1)^{k-1}.
\label{c12}
\eeq
Further, since the cavity function $y(r)\equiv g(r)e^{\beta \varphi(r)}$ must be continuous at $r=\la_1$ and $r=\la_2$, the two following conditions should also hold \cite{SYH12}:
\begin{subequations}
\label{3.5}
\beq
\frac{B_1}{S_3}=\left[e^{\beta(\epsilon_{1}-\epsilon_2)}-1\right]\sum_{\nu=1}^3\frac{s_\nu e^{(\la_1-1) s_\nu}}{S_1 +2S_2 s_\nu+3S_3 s_\nu^2} (A_0+B_0s_\nu),
\eeq
\beq
\frac{B_2}{S_3}=\left(e^{\beta\epsilon_{2}}-1\right)\sum_{\nu=1}^3\frac{s_\nu e^{(\la_2 -1)s_\nu}}{S_1 +2S_2 s_\nu+3S_3 s_\nu^2}
\left[A_0+B_0s_\nu+(A_1+B_1s_\nu)e^{-(\la_1-1) s_\nu}\right],
\label{3.6}
\eeq
\end{subequations}
where $s_\nu$ ($\nu=1,2,3$) are the three roots of the cubic equation $1+S_1 s+S_2 s^2+S_3 s^3=0$.

Equations \eqref{c9}--\eqref{3.5} still leave two parameters undetermined.
A simplifying assumption is that the coefficients $A_j$  ($j=0,1, 2$) may be fixed at their zero-density values, namely
\beq
A_0=e^{-\beta\epsilon_1},\quad A_1=e^{-\beta\epsilon_2}-e^{-\beta\epsilon_1},\quad A_2=1-e^{-\beta\epsilon_2}.
\label{3.2}
\eeq

This closes the problem of determining the nine parameters in terms of $\eta$, $\la_1$, $\la_2$, $\beta \epsilon_1$, and $\beta \epsilon_2$.  In fact, Equation \eqref{c9} allows us to express $S_1$, $S_2$, $S_3$, and $B_0$ as linear combinations of $B_1$ and $B_2$ so that in the end,  one only has to solve (numerically) the coupled transcendental Equation \eqref{3.5}. 

\begin{adjustwidth}{-\extralength}{0cm}

\reftitle{References}

 \PublishersNote{}
\end{adjustwidth}
\end{document}